\documentclass[prd,showpacs,showkeys,nofootinbib,twocolumn]{revtex4-1}

\usepackage{graphicx}
\usepackage{amsmath,amsfonts,amssymb,amsxtra,latexsym}

\usepackage{hyperref}
\hypersetup{
    colorlinks=true,
    linkcolor=red,
    citecolor=blue,      
    urlcolor=blue,
}

\usepackage[utf8]{inputenc}
\DeclareUnicodeCharacter{2212}{-}

\begin{document}

\title{Some Gravity Waves in Isotropic Cosmologies}

\author{Peter A. Hogan}
\email{peter.hogan@ucd.ie}
\affiliation{School of Physics, University College Dublin, Belfield, Dublin 4, Ireland} 

\author{Dirk Puetzfeld}
\email{dirk.puetzfeld@zarm.uni-bremen.de}
\homepage{http://puetzfeld.org}
\affiliation{University of Bremen, Center of Applied Space Technology and Microgravity (ZARM), 28359 Bremen, Germany} 

\date{ \today}

\begin{abstract}
We construct metric perturbations of two families of isotropic expanding universes describing gravitational waves propagating through these universes. The waves are non--planar and owe their wave front expansion solely to the expansion of the universes. The presence of this radiation leads to a small perturbation of the perfect fluid matter content of the universes by the appearance of an anisotropic stress. We then construct exact models of gravity waves in these universes. In this case the matter content of the models consists of the perfect fluid matter supplemented by anisotropic stress and lightlike matter traveling with the waves. Under appropriate conditions of approximation the lightlike matter can be neglected and the exact models coincide with the perturbative models.
\end{abstract}


\maketitle


\section{Introduction} \label{sec:introduction}

In an early study of waves in an expanding universe Schr\"odinger \cite{Schroedinger:1956:1} made the important observation that in such universes \emph{there are no plane waves filling the whole of space}. However the spatially homogeneous and isotropic Robertson--Walker geometry has the property that if the homogeneous hypersurfaces (the $T={\rm constant}$ hypersurfaces if $T$ is the cosmic time) have constant curvature $k=0$ or $-1$ then one can find null hypersurfaces in these spacetimes having the property that their 2--dimensional intersections with the homogeneous hypersurfaces are isometric to the Euclidean 2--plane. If such null hypersurfaces are the histories of wave fronts of gravitational waves propagating in these universes then these waves are arguably the simplest one can find in this context. Their wave fronts are expanding but the expansion is solely due to the expansion of the universe. The object of this paper is to construct perturbations of the Friedman--Lema\^itre cosmological models whose histories are these null hypersurfaces in the Robertson--Walker geometry and then to construct exact spacetime models of these waves.

There exists in the literature a two--fold approach to perturbations of cosmological models. In one approach perturbations of the Robertson--Walker line elements are used \cite{Bardeen:1982,Bardeen:1988}. This naturally entails a careful identification of terms arising which may or may not be transformed away by gauge transformations. Alternatively there exists the gauge invariant and covariant approach which involves working only with gauge invariant variables \cite{Ellis:Bruni:1989}. For perturbations of isotropic cosmologies these variables (which all vanish when specialized to isotropy) consist of the spatial gradients of the proper density, isotropic pressure and scalar expansion of the matter world lines, the 4--acceleration of the matter world lines, their shear and vorticity, the anisotropic stress and energy flux (or heat flow) of the matter and the so--called electric and magnetic parts of the Weyl conformal curvature tensor. The equations satisfied by these variables are obtained from the Ricci identities, the Bianchi identities and, for the consistency of the field equations and the Bianchi identities, the vanishing covariant divergence of the matter energy--momentum--stress tensor. Working in the context of this latter approach, perturbations describing gravitational waves have been obtained by looking for gauge invariant variables having an arbitrary dependence on a scalar function \cite{Hogan:Ellis:1997,Hogan:OShea:2002}. This was an implementation of Trautman's \cite{Trautman:1962} characterization of waves, namely, that \emph{waves can propagate information} with the information encoded in the arbitrary scalar function. 

The paper is organized as follows: In section \ref{sec:2} we describe the isotropic cosmologies and in doing so introduce our notations and sign conventions. In section \ref{sec:3} we construct a metric perturbation of the $k=0$ Robertson--Walker geometry and demonstrate how it describes gravitational waves propagating through this universe and how the presence of the waves distorts the isotropic perfect fluid matter distribution. This is followed in section \ref{sec:4} by the corresponding perturbation of the $k=-1$ Robertson--Walker geometry and its physical interpretation. In section \ref{sec:5} we outline how the metric perturbations of sections \ref{sec:3} and \ref{sec:4} result in the satisfaction of the equations of the gauge invariant and covariant perturbation theory. In section \ref{sec:6} we construct an exact model of gravity waves in the $k=0$ case and demonstrate that the resulting matter distribution consists of an isotropic perfect fluid, an anisotropic stress and lightlike matter traveling with the gravitational waves (i.e.\ having the same propagation direction in spacetime as the waves). The case of $k=-1$ corresponding to that described in section \ref{sec:6} is given in section \ref{sec:7}. The paper ends with a brief discussion of our results in section \ref{sec:discussion}.

\section{Geometrical Preliminaries}\label{sec:2}

Our starting point is the Friedman--Lema\^itre cosmological models with a spatially homogeneous and isotropic Robertson--Walker geometry. These are described by the metric tensor given via the line element
\begin{equation}\label{1}
ds^2=g_{ij}\,dX^i\,dX^j=\frac{S^2(T)(dX^2+dY^2+dZ^2)}{\left (1+\frac{k}{4}(X^2+Y^2+Z^2)\right )^2}-dT^2\ ,
\end{equation}
with $k=0, \pm 1$ and the scale factor $S(T)\geq 0$ for $T\geq 0$. The coordinates $X^i=(X, Y,Z, T)$ correspond to $i=1, 2, 3, 4$. We use units for which the gravitational constant $G=1$ and the speed of light in a vacuum $c=1$. The integral curves of the unit timelike vector field $u^i=\delta^i_4$ are the world lines of the constituent particles of the matter in the universe. The matter is necessarily a perfect fluid with energy--momentum--stress tensor
\begin{equation}\label{2}
T^{ij}=(\mu+p)u^i\,u^j+p\,g^{ij}\ .
\end{equation}
Here $g_{ij}\,u^i\,u^j=u_j\,u^j=-1$, $\mu$ is the proper--density of the fluid and $p$ is the isotropic pressure. An equation of state takes the form $p=p(\mu)$ but we will leave this unspecified throughout.

Our sign convention for the components $R_{ijkl}$ of the Riemann curvature tensor is fixed by writing the Ricci identities for an arbitrary vector field $v^i$ in the form
\begin{equation}\label{3}
v_{i;jk}-v_{i;kj}=v_l\,R^l{}_{ijk}\ ,
\end{equation}
with $v_i=g_{ij}\,v^j$ and the semicolon denotes covariant differentiation with respect to the Riemannian connection calculated with the metric tensor $g_{ij}$ while a comma will denote partial differentiation where appropriate. It is also useful to note the Ricci identities for any covariant tensor with components $A_{ij}$ now reads
\begin{equation}\label{4}
A_{ij;kl}-A_{ij;lk}=A_{pj}\,R^p{}_{ikl}+A_{ip}\,R^p{}_{jkl}\ .
\end{equation}
We write the Ricci tensor components as $R_{ij}=R^k{}_{ikj}$, the Ricci scalar is $R=g^{ij}\,R_{ij}$ and the Einstein tensor is $G_{ij}=R_{ij}-\frac{1}{2}\,g_{ij}\,R$. Einstein's field equations read
\begin{equation}\label{5}
G_{ij}=T_{ij}\ ,
\end{equation}
where we have absorbed a factor $8\,\pi$ into $T_{ij}$ on the right hand side here. We are thus following the sign conventions and practice of Ellis \cite{Ellis:1971}. Finally we note that the Weyl conformal curvature tensor components $C_{ijkl}$ are given by 
\begin{eqnarray}
C_{ijkl}&=&R_{ijkl}-\frac{1}{2}(g_{ik}\,R_{jl}+g_{jl}\,R_{ik}-g_{jk}\,R_{il}-g_{il}\,R_{jk})\nonumber\\
&&+\frac{R}{6}(g_{ik}\,g_{jl}-g_{il}\,g_{jk})\ .\label{5'}
\end{eqnarray}

For the special case of (\ref{1}) and (\ref{2}) the field equations (\ref{5}) read
\begin{equation}\label{6}
G_{ij}=\mu\,u_i\,u_j+p\,h_{ij}\ \ \ {\rm with}\ \ \ h_{ij}=g_{ij}+u_i\,u_j\ ,
\end{equation}
and $\mu=\mu(T)$, $p=p(T)$ since isotropy requires $h^i_j\,\mu_{,i}=0=h^i_j\,p_{,i}$ where $h^i_j=g^{ik}\,h_{kj}$ is the projection tensor which projects vectors orthogonal to $u^i$. For isotropy relative to any integral curve of $u^i$ there can be no preferred directions orthogonal to $u^i$ at any point of any integral curve. The field equations calculated with the metric tensor given by (\ref{1}) result in
\begin{equation}\label{7}
\mu=\frac{3\,\dot S^2}{S^2}+\frac{3\,k}{S^2}\ \ \ {\rm and}\ \ \ p=-\frac{\dot S^2}{S^2}-\frac{2\,\ddot S}{S}-\frac{k}{S^2}\ ,
\end{equation}
with a dot indicating differentiation with respect to $T$. The integral curves of $u^i$ are \emph{geodesic}
\begin{equation}\label{8}
\dot u^i\equiv u^i{}_{;j}\,u^j=0\ ,
\end{equation}
\emph{twist--free}
\begin{equation}\label{9}
\omega_{ij}=u_{[i;j]}+\dot u_{[i}\,u_{j]}=0\ ,
\end{equation}
with the square brackets denoting skew symmetrisation, \emph{shear--free}
\begin{equation}\label{10}
\sigma_{ij}=u_{(i;j)}+\dot u_{(i}\,u_{j)}-\frac{1}{3}\vartheta\,h_{ij}=0\ ,
\end{equation}
with round brackets denoting symmetrisation, and with \emph{expansion} 
\begin{equation}\label{11}
\vartheta=u^i{}_{;i}=3\,\frac{\dot S}{S}>0\ ,
\end{equation}
since $\dot S>0$ on account of the Hubble expansion of the universe. In addition these spacetimes are conformally flat and the Riemann curvature tensor is given by (\ref{5'}) and (\ref{6}) to read
\begin{eqnarray}
R_{ijkl}&=&\frac{1}{2}(\mu+p)\{g_{ik}\,u_j\,u_l+g_{jl}\,u_i\,u_k-g_{jk}\,u_i\,u_l\nonumber\\
&&-g_{il}\,u_j\,u_k\}+\frac{1}{3}\,\mu\,(g_{ik}\,g_{jl}-g_{il}\,g_{jk})\ .\label{11'}
\end{eqnarray}

\section{The case $k=0$}\label{sec:3}

For the case of $k=0$ a simple perturbation is given by a line element of the form
\begin{eqnarray}
ds^2&=&S^2\,[(1+2\,\alpha)dX^2+4\,\beta\,dX\,dY+(1-2\,\alpha)dY^2] \nonumber \\
&&+S^2dZ^2-dT^2\ ,\label{12}
\end{eqnarray}
where $\alpha=\alpha(X, Y, Z, T)$ and $\beta=\beta(X, Y, Z, T)$ are small of first order and we will consistently neglect second order and higher order terms. This form of perturbed line element has a number of useful properties. In the spacetime with line element (\ref{12}) the vector field $u^i=\delta^i_4$ is a unit timelike vector field which is geodesic, twist--free and has expansion $\vartheta$ and shear $\sigma_{ij}$ given by
\begin{equation}\label{13}
\vartheta=3\frac{\dot S}{S}\ \ \ {\rm and}\ \ \ \sigma_{ij}=u_{i;j}-\frac{1}{3}\vartheta\,h_{ij}=\Gamma^4_{ij}-\frac{\dot S}{S}\,h_{ij}\ ,
\end{equation}
respectively where $\Gamma^i_{jk}$ are the Christoffel symbols calculated with the metric tensor given via the line element (\ref{12}). Hence we find that $\sigma_{ij}=0$ except for
\begin{equation}\label{14}
\sigma_{11}=S^2\,\alpha_T\ ,\ \sigma_{22}=-S^2\,\alpha_T\ ,\ \sigma_{12}=S^2\,\beta_T\ ,
\end{equation}
with subscripts, here and below, denoting partial differentiation. This perturbed shear of the matter world lines is small of first order and can be written neatly as
\begin{equation}\label{15}
\sigma^{ij}=\bar\sigma\,m^i\,m^j+\sigma\,\bar m^i\,\bar m^j,
\end{equation}
with
\begin{equation}\label{16}
\sigma=\alpha_T+i\beta_T\ \ \ {\rm and}\ \ \ m^i\,\frac{\partial}{\partial X^i}=\frac{1}{S\,\sqrt{2}}\left (\frac{\partial}{\partial X}+i\,\frac{\partial}{\partial Y}\right )\ ,
\end{equation}
with the bar denoting complex conjugation. For the histories of gravitational waves in the spacetime with line element (\ref{12}) we take the null hypersurfaces
\begin{equation}\label{17}
\phi(Z, T)=Z-\tau(T)={\rm constant},\ \ \ {\rm with}\ \ \ \frac{d\tau}{dT}=\frac{1}{S}\ .
\end{equation}
We see that 
\begin{equation}
\phi_{,i}=(0, 0, 1, -S^{-1})
\end{equation}
and
\begin{equation}\label{18}
g^{ij}\,\phi_{,j}=(0, 0, S^{-2}, S^{-1})\ \ \Rightarrow\ \ g^{ij}\,\phi_{,i}\,\phi_{,j}=0\ .
\end{equation}
Next we require $\alpha, \beta$ in (\ref{12}) to satisfy the Cauchy--Riemann equations
\begin{equation}\label{19}
\alpha_X=-\beta_Y\ \ \ {\rm and}\ \ \ \alpha_Y=\beta_X\ .
\end{equation}
This will ensure that the intersections of the null hypersurfaces $\phi={\rm constant}$ with the spacelike hypersurfaces $T={\rm constant}$ are isometric to Euclidean 2--space. From (\ref{12}) the induced line elements on these 2--surfaces are given, modulo a multiplicative constant, by
\begin{equation}\label{20}
dl^2=(1+2\,\alpha)\,dX^2+4\,\beta\,dX\,dY+(1-2\,\alpha)\,dY^2\ .
\end{equation}
The Gaussian curvature of such 2--surfaces is 
\begin{equation}\label{21}
K=\alpha_{XX}-\alpha_{YY}+2\,\beta_{XY}\ ,
\end{equation}
and $K=0$ on account of the assumption (\ref{19}). The final simplification, before examining the field equations, is to require the null hypersurfaces (\ref{17}) to be shear--free in the optical sense. To achieve this we first calculate the covariant derivative of $\phi_{,i}$ with respect to the Riemannian connection calculated with the metric given by the line element (\ref{12}), neglecting second order and smaller terms. We find 
\begin{eqnarray}
\phi_{,i;j}&=&\frac{\dot S}{S^2}\,g_{ij}+\frac{\dot S}{S}\,(u_i\,\phi_{,j}+u_j\,\phi_{,i})\nonumber\\
&&+\frac{1}{S^2}\{\alpha_Z+S\,\alpha_T-i\,(\beta_Z+S\,\beta_T)\}\,m_i\,m_j\nonumber\\
&&+\frac{1}{S^2}\{\alpha_Z+S\,\alpha_T+i\,(\beta_Z+S\,\beta_T)\}\,\bar m_i\,\bar m_j.\label{22}
\end{eqnarray}
The right hand side of this equation has the correct algebraic form to ensure that $\phi={\rm constant}$ are shear--free in the optical sense \cite{Robinson:Trautman:1983} provided
\begin{eqnarray}
&& \alpha_Z+S\,\alpha_T+i\,(\beta_Z+S\,\beta_T)=0 \nonumber \\
&\Rightarrow& \alpha=\alpha(X, Y, \phi)\ \ {\rm and}\ \ \beta=\beta(X, Y, \phi)\ .\label{23}
\end{eqnarray}
This result can also be seen by first taking $m^i$ in (\ref{16}) in the more accurate form
\begin{eqnarray}
m^i\,\frac{\partial}{\partial X^i}&=&\frac{1}{S\,\sqrt{2}}\left\{ (1-\alpha-i\,\beta)\,\frac{\partial}{\partial X} \right. \nonumber \\
&&\left.+i\,(1+\alpha+i\,\beta)\,\frac{\partial}{\partial Y}\right \}\ ,\label{23'}
\end{eqnarray}
and then finding that 
\begin{equation}\label{23''}
\phi_{,i;j}\,m^i\,m^j=\frac{1}{S^2}\{\alpha_Z+S\,\alpha_T+i(\beta_Z+S\,\beta_T)\}\ .
\end{equation}
We see from (\ref{22}) that $\phi={\rm constant}$ are generated by null geodesics having expansion (in the optical sense) 
\begin{equation}\label{24}
\frac{1}{2}\phi_{,i}{}^{;i}=\frac{\dot S}{S^2}\ .
\end{equation}
As a consequence of (\ref{23}) we can write $\bar\sigma$, with $\sigma$ given by (\ref{16}), as
\begin{equation}\label{25}
\bar\sigma=-\frac{(\alpha'-i\,\beta')}{S}\ ,
\end{equation}
with the prime denoting differentiation with respect to $\phi$. If we put $\zeta=X+i\,Y$ then the Cauchy--Riemann equations (\ref{19}) mean that we can write $\alpha'-i\,\beta'={\cal G}(\zeta, \phi)$ and so we have
\begin{equation}\label{26}
\bar\sigma=-\frac{{\cal G}(\zeta, \phi)}{S(T)}\ .
\end{equation}
The components $G_{ij}$ of the Einstein tensor calculated with the metric tensor given by (\ref{12}), along with the extra conditions (\ref{19}) and (\ref{23}), results in $G_{ij}=0$ except for
\begin{eqnarray}
G_{11}&=&-\dot S^2-2\,S\,\ddot S-(2\,\dot S^2+4\,S\,\ddot S)\,\alpha-2\,\dot S\,\alpha'\ ,\label{27}\\
G_{22}&=&-\dot S^2-2\,S\,\ddot S+(2\,\dot S^2+4\,S\,\ddot S)\,\alpha+2\,\dot S\,\alpha'\ ,\label{28}\\
G_{12}&=&-(2\,\dot S^2+4\,S\,\ddot S)\,\beta-2\,\dot S\,\beta'\ ,\label{29}\\
G_{33}&=&-\dot S^2-2\,S\,\ddot S\ ,\label{30}\\
G_{44}&=&3\,\frac{\dot S^2}{S^2}\ .\label{31}
\end{eqnarray}
Thus Einstein's equations (\ref{5}) give us a perturbed matter distribution with energy--momentum--stress tensor
\begin{equation}\label{32}
T_{ij}=(\mu+p)\,u_i\,u_j+p\,g_{ij}+\pi_{ij}\ ,
\end{equation}
which is a perturbation of the perfect fluid isotropic matter distribution by the addition of an anisotropic stress $\pi_{ij}=\pi_{ji}$ with $\pi_{ij}\,u^j=0$ and $g^{ij}\,\pi_{ij}=0$. Reading off from (\ref{27})--(\ref{31}) and (\ref{32}) we find that
\begin{equation}\label{33}
\mu=3\,\frac{\dot S^2}{S^2}\ \ \ {\rm and}\ \ \ p=-\frac{\dot S^2}{S^2}-2\,\frac{\ddot S}{S}\ ,
\end{equation}
which shows that these quantities are unperturbed from their isotropic values, while $\pi_{ij}=0$ except for 
\begin{equation}\label{34}
\pi_{11}=-2\,\dot S\,\alpha'\ ,\ \pi_{12}=-2\,\dot S\,\beta'\ ,\ \pi_{22}=2\,\dot S\,\alpha'\ .
\end{equation} 
The latter can be written neatly as
\begin{equation}\label{35}
\pi^{ij}=\pi\,\bar m^i\,\bar m^j+\pi\,m^i\,m^j\ ,
\end{equation}
with
\begin{equation}\label{36}
\bar\pi=-2\,\frac{\dot S}{S^2}\,{\cal G}(\zeta, \phi)=-\phi_{,i}{}^{;i}\,{\cal G}\ ,
\end{equation}
using (\ref{24}). Thus we see explicitly that the existence of the anisotropic stress in the matter distribution has a geometrical origin in the necessary expansion of the wave fronts due to the expansion of the universe.

The perturbed Weyl conformal curvature tensor $C_{ijkl}$ can be given equivalently in terms of its \emph{electric part}
\begin{equation}\label{37}
E_{ik}=C_{ijkl}\,u^j\,u^l\ ,
\end{equation}
and its \emph{magnetic part}
\begin{equation}\label{38}
H_{ik}=C^*_{ijkl}\,u^j\,u^l\ \ \ {\rm with}\ \ \ C^*_{ijkl}=\frac{1}{2}\,\eta_{ijpq}\,C^{pq}{}_{kl}\ ,
\end{equation}
and here $\eta_{ijkl}=\sqrt{-g}\,\epsilon_{ijkl}$ with $g={\rm det}(g_{ij})$ and $\epsilon_{ijkl}$ is the four dimensional Levi--Civita permutation symbol for which we take $\epsilon_{1234}=+1$. For the model we have constructed we find
\begin{eqnarray}
E^{11}+i\,H^{11}&=&-\frac{{\cal G}'}{S^4}\ ,\nonumber \\ 
E^{12}+i\,H^{12}&=&-i\,\frac{{\cal G}'}{S^4}\ ,\nonumber \\
E^{22}+i\,H^{22}&=&\frac{{\cal G}'}{S^4}\ ,\label{39}
\end{eqnarray}
with all remaining components of $E^{ij}+i\,H^{ij}$ vanishing. In terms of the complex vector field $m^i$ given in (\ref{16}) we can write this as
\begin{equation}\label{40}
E^{ij}+i\,H^{ij}=-2\frac{{\cal G}'}{S^2}\,m^i\,m^j\ .
\end{equation}
We note that ${\cal G}={\cal G}(\zeta, \phi)$ and ${\cal G}'=\partial{\cal G}/\partial\phi$. We pass from the electric and magnetic parts of the Weyl tensor to the Weyl tensor itself using \cite{Ellis:1971}
\begin{eqnarray}
C_{ijkl}+i\,C^*_{ijkl}&=&w_{ijpq}\,w_{klrs}\,u^p\,u^r\,(E^{qs}+i\,H^{qs}) \nonumber \\
&=&-2\frac{{\cal G}'}{S^2}\,w_{ijpq}\,w_{klrs}\,u^p\,u^r\,m^q\,m^s\ ,\label{41}
\end{eqnarray}
with
\begin{equation}\label{42}
w_{ijpq}=g_{ip}\,g_{jq}-g_{iq}\,g_{jp}+i\,\eta_{ijpq}\ ,
\end{equation}
we find that 
\begin{equation}\label{43}
(C_{ijkl}+C^*_{ijkl})\,\phi^{,l}=0\ ,
\end{equation}
(with $\phi^{,l}=g^{lk}\,\phi_{,k}$) on account of 
\begin{equation}\label{44}
w_{klrs}\,\phi^{,l}\,u^r\,m^s=\frac{1}{\sqrt{2}}(\delta^1_k+i\,\delta^2_k)\,(1-\epsilon_{1234})=0\ .
\end{equation}
The algebraic condition (\ref{43}) confirms that the perturbed field described by (\ref{40}) is purely radiative with 
propagation direction in spacetime $\phi^{,i}$ since this corresponds to a Petrov type N field with degenerate principal null direction $\phi^{,i}$.

We have presented the perturbed shear, anisotropic stress and electric and magnetic parts of the Weyl tensor in the specific forms involving (\ref{26}), (\ref{36}) and (\ref{40}) to facilitate comparison with the more general gauge invariant perturbations derived in \cite{Hogan:OShea:2002}. 

\section{The case $k=-1$}\label{sec:4}

Our starting point now is the isotropic cosmological model described by the line element (\ref{1}) with $k=-1$. To put this line element into a suitable form for our purposes we make the coordinate transformation
\begin{equation}\label{45}
X=8\,x\,\lambda^{-1}\ ,\ Y=8\,y\,\lambda^{-1}\ ,\ Z=-2-8\,(z'-2)\,\lambda^{-1}\ ,
\end{equation}
with $\lambda=x^2+y^2+(z'-2)^2$. This results in (\ref{1}) with $k=-1$ taking the form
\begin{equation}\label{46}
ds^2=S^2(T)\,z'^{-2}\,(dx^2+dy^2+dz'^2)-dT^2\ .
\end{equation}
Now put $z'=e^{-z}$ and we arrive at
\begin{equation}\label{47}
ds^2=S^2(T)\,e^{2\,z}\,(dx^2+dy^2)+S^2(T)\,dz^2-dT^2\ .
\end{equation}
The perturbed line element of interest to us in this case reads
\begin{eqnarray}
ds^2&=&S^2\,e^{2\,z}[(1+2\,\alpha)dx^2+4\,\beta\,dx\,dy+(1-2\,\alpha)dy^2]\nonumber \\
&&+S^2dz^2-dT^2\ ,\label{48}
\end{eqnarray}
where $\alpha=\alpha(x, y, z, T)$ and $\beta=\beta(x, y, z, T)$ are small of first order as before. The null hypersurfaces of interest are
\begin{equation}\label{49}
\phi(z, T)=z-\tau(T)={\rm constant}\ \ \ {\rm with}\ \ \ \frac{d\tau}{dT}=\frac{1}{S}\ .
\end{equation}
The intersections of $\phi={\rm constant}$ and $T={\rm constant}$ are isometric to Euclidean 2--space if $\alpha, \beta$ satisfy the Cauchy--Riemann equations 
\begin{equation}\label{50}
\alpha_x=-\beta_y\ \ \ {\rm and}\ \ \ \alpha_y=\beta_x\ .
\end{equation}
Using the complex null vector field
\begin{eqnarray}
m^i\,\frac{\partial}{\partial x^i}&=&\frac{e^{-z}}{S\,\sqrt{2}}\left\{ (1-\alpha-i\,\beta)\,\frac{\partial}{\partial x} \right.\nonumber\\
&&\left.+i\,(1+\alpha+i\,\beta)\,\frac{\partial}{\partial y}\right \}\ ,\label{51}
\end{eqnarray}
we have
\begin{equation}\label{52}
\phi_{,i;j}\,m^i\,m^j=\frac{1}{S^2}\{\alpha_z+S\,\alpha_T+i\,(\beta_z+S\,\beta_T)\}\ ,
\end{equation}
and thus the null hypersurfaces $\phi={\rm constant}$ are shear--free in the optical sense if
$\alpha=\alpha(x, y, \phi)$ and $\beta=\beta(x, y, \phi)$. The perturbed shear of the matter world lines, given by (\ref{13}) but applied to the present $k=-1$ case, gives $\sigma_{ij}=0$ except for 
\begin{equation}\label{53}
\sigma_{11}=S^2e^{2\,z}\,\alpha_T=-\sigma_{22}\\ \ {\rm and}\ \ \ \sigma_{12}=S^2e^{2\,z}\,\beta_T\ .
\end{equation}
With $m^i$ given by (\ref{51}) we can write this as
\begin{equation}\label{54}
\sigma^{ij}=\bar\sigma\,m^i\,m^j+\sigma\,\bar m^i\,\bar m^j\ ,
\end{equation}
with
\begin{equation}\label{55}
\sigma=\alpha_T+i\,\beta_T\ .
\end{equation}
For later comparison with the gauge invariant approach to this modelling we put this in the form
\begin{equation}\label{56}
\bar\sigma=\alpha_T-i\,\beta_T=-\frac{1}{S}(\alpha'-i\,\beta')=-\frac{e^{-z}\,{\cal G}}{S(T)}\ ,
\end{equation}
with
\begin{equation}\label{57}
{\cal G}(x, y, z, T)=e^{z}\,(\alpha'-i\beta')=a(\zeta, \phi)\,e^{\frac{1}{2}(z+\tau)}\ ,
\end{equation}
$a(\zeta, \phi)=(\alpha'-i\,\beta')\exp{\{\frac{1}{2}\phi\}}$ and $\zeta=x+i\,y$. We note that with $D=\partial/\partial z+S\,\partial/\partial T$ the complex valued function ${\cal G}$ here satisfies
\begin{equation}\label{58}
D{\cal G}={\cal G}\ \ \ \Rightarrow\ \ \ D^2{\cal G}-{\cal G}=0\ .
\end{equation}

The non--vanishing components of the Einstein tensor in this case (neglecting second order and smaller terms in $\alpha$ and $\beta$) are
\begin{eqnarray}
G_{11}&=&e^{2\,z}S^2p-2\,e^{2\,z} (1+\dot S)\,\alpha'\ ,\label{59}\\
G_{22}&=&e^{2\,z}S^2p+2\,e^{2\,z} (1+\dot S)\,\alpha'\ ,\label{60}\\
G_{12}&=&2\,e^{2\,z}S^2\,\beta\,p-2\,e^{2\,z}(1+\dot S)\,\beta'\ ,\label{61}\\
G_{33}&=&S^2p\ ,\label{62}\\
G_{44}&=&\mu\ ,\label{63}
\end{eqnarray}
with
\begin{equation}\label{64}
p=\frac{1}{S^2}-\frac{\dot S^2}{S^2}-\frac{2\,\ddot S}{S}\ \ \ {\rm and}\ \ \ \mu=\frac{3\,(\dot S^2-1)}{S^2}\ .
\end{equation}
With this Einstein tensor the right hand side of the field equations is a energy--momentum--stress tensor of the form (\ref{32}) with $u_i=-\delta^4_i$, $p$ and $\mu$ given by (\ref{64}) (the unperturbed isotropic expressions in this case) and the anisotropic stress components $\pi_{ij}$ vanish with the exception of
\begin{equation}
\pi_{11}=-\pi_{22}=-2\,e^{2\,z}(1+\dot S)\,\alpha' \nonumber \\
\end{equation} 
and
\begin{equation}\label{65}
\pi_{12}=\pi_{21}=-2\,e^{2\,z}(1+\dot S)\,\beta'\ .
\end{equation} 
Using the complex null vector field (\ref{51}) and the function ${\cal G}$ given by (\ref{57}) we can write the anisotropic stress tensor in the form
\begin{equation}\label{66}
\pi^{ij}=\bar\pi\,m^i\,m^j+\pi\,\bar m^i\,\bar m^j\ ,
\end{equation}
with
\begin{equation}\label{67}
\bar\pi=-\frac{2}{S^2}(1+\dot S)(\alpha'-i\,\beta')=-\frac{2\,e^{-z}}{S^2}(1+\dot S)\,{\cal G}\ .
\end{equation}
We note that with $\phi$ given by (\ref{49}) and ${\cal G}$ satisfying (\ref{58}) we can write
\begin{equation}\label{68}
\bar\pi=-\frac{2\,e^{-z}}{S^2}(D{\cal G}+\dot S\,{\cal G})=-\phi_{,i}{}^{;i}\,e^{-z}\,{\cal G}\ .
\end{equation}
We see from this that the anisotropic stress is present on account of the expansion of the wave fronts.

The non--vanishing perturbed Weyl tensor components in this case read
\begin{eqnarray}
C_{1313}&=&-S\,C_{1314}=S^2C_{1414}=-S^2e^{2\,z}\,(\alpha'+\alpha''),\label{69a} \\
C_{2323}&=&-S\,C_{2324}=S^2C_{2424}=S^2e^{2\,z}\,(\alpha'+\alpha''),\label{70a}
\end{eqnarray}
and
\begin{eqnarray}
C_{1323}&=&-S\,C_{1324}=-S\,C_{1423}=S^2C_{1424} \nonumber \\
&=&-S^2e^{2\,z}\,(\beta'+\beta'')\ .\label{71a}
\end{eqnarray}
The components of the dual of the Weyl tensor are now found to be
\begin{eqnarray}
C^*_{1313}&=&-S\,C^*_{1314}=S^2C^*_{1414}=S^2e^{2\,z}\,(\beta'+\beta''),\label{72a}\\
C^*_{2323}&=&-S\,C^*_{2324}=S^2C^*_{2424}=-S^2e^{2\,z}\,(\beta'+\beta''), \label{73a}
\end{eqnarray}
and
\begin{eqnarray}
C^*_{1323}&=&-S\,C^*_{1324}=-S\,C^*_{2314}=S^2C^*_{1424} \nonumber \\
&=&-S^2e^{2\,z}\,(\alpha'+\alpha'')\ . \label{74a}
\end{eqnarray}
With $\phi(z, T)$ given by (\ref{49}) we see that $C_{ijkl}$ and $C^*_{ijkl}$ satisfy
\begin{equation}\label{75a}
C_{ijkl}\,\phi^{,l}=0=C^*_{ijkl}\,\phi^{,l}\ ,
\end{equation}
which demonstrates that the perturbed Weyl tensor is purely radiative with propagation direction in spacetime $\phi^{,i}$ and the histories of the wave fronts in spacetime are the null hypersurfaces $\phi={\rm constant}$. In addition the electric part of the Weyl tensor $E^{ij}$ has the following non--vanishing components
\begin{equation}
E^{11}=-E^{22}=-S^{-4}\,e^{-2\,z}\,(\alpha'+\alpha'')\nonumber 
\end{equation}
and
\begin{equation}
E^{12}=-S^{-4}\,e^{-2\,z}\,(\beta'+\beta'')\ ,\label{76a}
\end{equation}
while the magnetic part of the Weyl tensor has the non--vanishing components
\begin{equation}
H^{11}=-H^{22}=S^{-4}\,e^{-2\,z}\,(\beta'+\beta'') \nonumber
\end{equation}
and
\begin{equation}\label{77a}
H^{12}=-S^{-4}\,e^{-2\,z}\,(\alpha'+\alpha'')\ .
\end{equation}
In terms of the complex null vector field (\ref{52}) and the complex valued function ${\cal G}$ in (\ref{56}) we can summarize these tensor fields as
\begin{eqnarray}
E^{ij}+i\,H^{ij}&=&-\frac{2}{S^2}\{\alpha'+\alpha''-i\,(\beta'+\beta'')\}\,m^i\,m^j \nonumber \\
&=&-\frac{2\,e^{-z}}{S^2}\,\frac{\partial {\cal G}}{\partial z}\,m^i\,m^j\ .\label{78a}
\end{eqnarray}
The perturbed shear, anisotropic stress and electric and magnetic parts of the Weyl tensor are presented here using the forms given by (\ref{56}), (\ref{67}) and (\ref{78a}) to make easy comparison with the more general gauge invariant perturbations obtained in \cite{Hogan:OShea:2002}.

\section{Gauge Invariant Perturbation Theory}\label{sec:5}

The non--vanishing gauge invariant perturbation variables above are $\sigma^{ab}$, $\pi^{ab}$, $E^{ab}$ and $H^{ab}$. These must satisfy \cite{Hogan:OShea:2002}
\begin{eqnarray}
E^{ab}&=&\frac{1}{2}\,\pi^{ab}-\dot\sigma^{ab}-\frac{2}{3}\,\vartheta\,\sigma^{ab}\ ,\label{69}\\
H^{ab}&=&-\sigma^{(a}{}_{g;c}\,\eta^{b)fgc}\,u_f\ ,\label{70}\\
\sigma^{ab}{}_{;b}&=&0\ ,\label{71}
\end{eqnarray}
which follow from the Ricci identities and
\begin{equation}\label{72}
\pi^{ab}{}_{;b}=0\ ,\end{equation}
which is a consequence of the vanishing divergence of the energy--momentum--stress tensor. In addition the Bianchi identities require
\begin{eqnarray}
\dot E^{bt}&+&\vartheta\,E^{bt}+H^{(b}{}_{s;d}\,\eta^{t)rsd}\,u_r \nonumber \\
&=&-\frac{1}{2}(\mu+p)\,\sigma^{bt}-\frac{1}{2}\dot\pi^{bt}-\frac{1}{6}\,\vartheta\,\pi^{bt}\ ,\label{73}
\end{eqnarray}
and
\begin{equation}\label{74}
\dot H^{bt}+\vartheta\,H^{bt}-E^{(b}{}_{s;d}\,\eta^{t)rsd}\,u_r=-\frac{1}{2}\eta^{(b}{}_{rad}\,\pi^{t)a;d}\,u^r\ .
\end{equation}
Once these equations are satisfied $\sigma^{bt}$ automatically satisfies the wave equation
\begin{eqnarray}
\sigma^{bt;d}{}_{;d}&-&\frac{2}{3}\,\vartheta\,\dot\sigma^{bt}+\left\{-\frac{1}{3}\,\dot\vartheta-\frac{4}{9}\,\vartheta^2+p-\frac{1}{3}\,\mu\right\}\,\sigma^{bt} \nonumber \\
&=&-\dot\pi^{bt}-\frac{2}{3}\,\vartheta\,\pi^{bt}\ . \label{75}
\end{eqnarray}
This is derived by substituting $E^{ab}$ and $H^{ab}$ from (\ref{69}) and (\ref{70}) into (\ref{73}) and, in particular, using
\begin{equation}\label{76}
\sigma^{td;b}{}_{;d}=\left (\frac{5}{6}\,\mu-\frac{1}{2}\,p\right )\,\sigma^{bt}\ ,
\end{equation}
which is obtained by contracting the Ricci identities (\ref{4}) with $A_{ij}=\sigma_{ij}$ and using the isotropic Riemann tensor (\ref{11'}).

To facilitate verification that (\ref{69})--(\ref{74}) are satisfied in the case of $k=-1$ above we note that the covariant derivatives here operate on first order small quantities and thus involve the Riemannian connection calculated with the metric of the isotropic background (\ref{47}). The non--vanishing connection components, in coordinates $x^i=(x, y, z, T)$ for $i=1, 2, 3, 4$ read
\begin{eqnarray}
\Gamma^1_{14}&=&\Gamma^2_{24}=\Gamma^3_{34}=\frac{\dot S}{S}\ ,\nonumber \\
\Gamma^1_{13}&=&\Gamma^2_{23}=1\ , \quad \Gamma^3_{11}=\Gamma^3_{22}=-e^{2\,z}\ ,\nonumber\\
\Gamma^4_{11}&=&\Gamma^4_{22}=e^{2\,z}\,\Gamma^4_{33}=S\,\dot S\,e^{2\,z}\ ,\label{77} 
\end{eqnarray}
and it is helpful to note that
\begin{eqnarray}
E^{1}{}_{s;d}\,\eta^{1rsd}\,u_r&=&-E^{2}{}_{s;d}\,\eta^{2rsd}\,u_r \nonumber \\
&=&-\frac{e^{-2\,z}}{S^5}\,(\beta'''+2\,\beta''+\beta')\ ,\label{78}\\
E^{(1}{}_{s;d}\,\eta^{2)rsd}\,u_r&=&\frac{e^{-2\,z}}{S^5}\,(\alpha'''+2\,\alpha''+\alpha')\ ,\label{79}\\
H^{1}{}_{s;d}\,\eta^{1rsd}\,u_r&=&-H^{2}{}_{s;d}\,\eta^{2rsd}\,u_r \nonumber \\
&=&-\frac{e^{-2\,z}}{S^5}\,(\alpha'''+2\,\alpha''+\alpha')\ , \label{80}\\
H^{(1}{}_{s;d}\,\eta^{2)rsd}\,u_r&=&-\frac{e^{-2\,z}}{S^5}\,(\beta'''+2\,\beta''+\beta')\ ,\label{81}
\end{eqnarray}
and
\begin{eqnarray}
\eta^{1}{}_{rad}\,\pi^{1a;d}\,u^r&=&-\eta^{2}{}_{rad}\,\pi^{2a;d}\,u^r \nonumber \\
&=&-2\,\frac{(1+\dot S)}{S^5}\,e^{-2\,z}\,(\beta''+\beta')\ ,\label{82}\\
\eta^{(1}{}_{rad}\,\pi^{2)a;d}\,u^r&=&2\,\frac{(1+\dot S)}{S^5}\,e^{-2\,z}\,(\alpha''+\alpha')\ .\label{83}
\end{eqnarray} 
For the case $k=0$ the covariant derivatives are with respect to the Riemannian connection calculated with the metric of the isotropic background (\ref{1}) (with $k=0$). In the coordinates $X^i=(X, Y, Z, T)$ for $i=1, 2, 3, 4$ the non--vanishing connection components are
\begin{equation}\label{84}
\Gamma^1_{14}=\Gamma^2_{24}=\Gamma^3_{34}=\frac{\dot S}{S}\ ,\ \Gamma^4_{11}=\Gamma^4_{22}=\Gamma^4_{33}=S\,\dot S\ .
\end{equation} 
The helpful formulas in this case read
\begin{eqnarray}
E^{1}_{s;d}\,\eta^{1rsd}\,u_r&=&-E^{2}_{s;d}\,\eta^{2rsd}\,u_r=-\frac{\beta''}{S^5}\ ,\label{85}\\
E^{(1}{}_{s;d}\,\eta^{2)rsd}\,u_r&=&\frac{\alpha''}{S^5}\ ,\label{86}\\
H^{1}{}_{s;d}\,\eta^{1rsd}\,u_r&=&-H^{2}{}_{s;d}\,\eta^{2rsd}\,u_r=-\frac{\alpha'''}{S^5}\ ,\label{87}\\
H^{(1}{}_{s;d}\,\eta^{2)rsd}\,u_r&=&-\frac{\beta'''}{S^5}\ ,\label{88}
\end{eqnarray}
and
\begin{eqnarray}
\eta^{1}{}_{rad}\,\pi^{1a;d}\,u^r&=&-\eta^{2}{}_{rad}\,\pi^{2a;d}\,u^r=-\frac{2\,\dot S}{S^5}\,\beta''\ ,\label{89}\\
\eta^{(1}{}_{rad}\,\pi^{2)a;d}\,u^r&=&\frac{2\,\dot S}{S^5}\,\alpha''\ .\label{90}
\end{eqnarray}

To accommodate the anisotropic stress in an otherwise perfect fluid matter distribution the world lines of the fluid particles must acquire a small amount of shear $\sigma_{ij}$. This follows from the fact that in both of the cases $k=0$ and $k=-1$ above the anisotropic stress satisfies
\begin{equation}\label{91}
\pi^{ij}=S\,\phi^{,k}{}_{;k}\,\sigma^{ij}\ .
\end{equation} 
Since the proper density $\mu$ and the isotropic pressure $p$ satisfy
\begin{equation}\label{92}
\mu+p=\frac{2}{S^2}(\dot S^2-S\,\ddot S+k)\ \ \ {\rm for}\ \ \ k=0, \pm 1\ ,
\end{equation}
we can conclude that if $\mu+p\neq 0$ then $\phi^{,k}{}_{;k}\neq 0$ for $k=0, -1$ and the anisotropic stress is solely due to the perturbed shear of the matter world lines. We note that in the isotropic model $\mu+p\neq0$ is a reasonable assumption as it ensures that the matter energy--momentum--stress tensor has a unique timelike eigenvector.

\section{An Exact Model When $k=0$}\label{sec:6}

Consider the line element, in coordinates $x^i=(x, y, T, \phi)$ for $i=1, 2, 3, 4$,
\begin{eqnarray}
ds^2=g_{ij}\,dx^i\,dx^j&=&S^2\{(dx-a\,d\phi)^2+(dy-b\,d\phi)^2\} \nonumber \\
&&+2\,S\,dT\,d\phi+S^2d\phi^2\ , \label{93}
\end{eqnarray}
with $S=S(T), a=a(x, y, \phi), b=b(x, y, \phi)$ and
\begin{equation}\label{94}
a_x=-b_y\ \ \ {\rm and}\ \ \ a_y=b_x\ ,
\end{equation}
with the subscripts denoting partial derivatives. If $a, b$ are small of first order this line element can be transformed into (\ref{12}) if only terms of first order are retained. 

To see this we write the first coefficient of $S^2$ in (\ref{12}) in the form
\begin{eqnarray}
&&(1+2\,\alpha)\,dX^2+4\,\beta\,dX\,dY+(1-2\,\alpha)\,dY^2\nonumber\\
&&=\{(1+\alpha)\,dX+\beta\,dY\}^2+\{\beta\,dX+(1-\alpha)\,dY\}^2\ ,\nonumber\\ \label{xr1}
\end{eqnarray}
which holds if we neglect second order terms in $\alpha$ and $\beta$. It follows from (\ref{19}) that we have two real--valued harmonic functions $f(X, Y, \phi)$ and $g(X, Y, \phi)$ such that
\begin{equation}\label{xr2}
\alpha=\frac{\partial f}{\partial X}\ ,\ \beta=\frac{\partial f}{\partial Y}\ \ {\rm and}\ \ \alpha=-\frac{\partial g}{\partial Y}\ ,\ \beta=\frac{\partial g}{\partial X}\ .
\end{equation}
Consequently we have
\begin{eqnarray}
(1+\alpha)\,dX+\beta\,dY&=&dX+\frac{\partial f}{\partial X}\,dX+\frac{\partial f}{\partial Y}\,dY\nonumber \\
&=&d(X+f)-f'\,d\phi\ ,\label{xr3}
\end{eqnarray}
and
\begin{eqnarray}
\beta\,dX+(1-\alpha)\,dY&=&dY+\frac{\partial g}{\partial Y}\,dY+\frac{\partial g}{\partial X}\,dX\nonumber\\
&=&d(Y+g)-g'\,d\phi\ ,\label{xr4}
\end{eqnarray}
with the prime as always denoting partial differentiation with respect to $\phi$. If we make the infinitesimal coordinate transformations
\begin{equation}\label{xr5}
x=X+f\ \ \ {\rm and}\ \ \ y=Y+g\ ,
\end{equation}
then (\ref{xr3}) and (\ref{xr4}) take the form
\begin{equation}\label{xr6}
(1+\alpha)\,dX+\beta\,dY=dx-a(x, y, \phi)\,d\phi\ ,
\end{equation}
and 
\begin{equation}\label{xr7}
\beta\,dX+(1-\alpha)\,dX=dy-b(x, y, \phi)\,d\phi\ ,
\end{equation}
respectively, remembering that we are neglecting second order small quantities. Here $a=f'$ and $b=g'$ and it follows from (\ref{xr2}) that to first order $a$ and $b$ satisfy (\ref{94}).

The 4--velocity of matter is given by
\begin{eqnarray}
u_i\,dx^i&=&-dT\ \nonumber \\
 \Leftrightarrow u^i\,\frac{\partial}{\partial x^i}&=&-\frac{a}{S}\,\frac{\partial}{\partial x}-\frac{b}{S}\,\frac{\partial}{\partial y}+\frac{\partial}{\partial T}-\frac{1}{S}\,\frac{\partial}{\partial\phi}\ ,\label{94a}
\end{eqnarray}
and thus
\begin{equation}\label{94b}
g_{ij}\,u^i\,u^j=u_j\,u^j=-1\ \ \ {\rm and}\ \ \ u_{i;j}=\Gamma^3_{ij}=u_{j;i}\ ,
\end{equation}
where the semicolon denotes covariant differentiation with respect to the Riemannian connection calculated with the metric given by (\ref{93}) and $\Gamma^3_{ij}$ are components of the Riemannian connection. Since from (\ref{93}) ${\rm det}(g_{ij})=-S^6$ we have exactly $\theta=u^i{}_{;i}=3\,\dot S/S$ and it follows from (\ref{94b}) that $\dot u_i=u_{i;j}\,u^j=0$ and $\omega_{ij}=0$ so that the integral curves of the vector field $u^i$ are expanding, geodesic and twist--free. The shear of these curves is given by
\begin{equation}\label{94c}
\sigma_{ij}=\Gamma^3_{ij}-\frac{\dot S}{S}\,(g_{ij}+\delta^3_i\,\delta^3_j)\ .
\end{equation}
The non--vanishing Christoffel symbols appearing in (\ref{94c}) are 
\begin{eqnarray}
\Gamma^3_{11}&=&S\,\dot S-a_x\,S\ ,\ \Gamma^3_{22}=S\,\dot S-b_y\,S\ ,\nonumber \\
\Gamma^3_{12}&=&-a_y\,S\ , \quad \Gamma^3_{33}=S^{-1}\dot S\ ,\nonumber\\
\Gamma^3_{14}&=&-a\,S\,\dot S+(a\,a_x+b\,b_x)\,S\ ,\nonumber \\
\Gamma^3_{24}&=&-b\,S\,\dot S+(a\,a_y+b\,b_y)\,S\ ,\ \Gamma^3_{34}=\dot S\ ,\nonumber\\
\Gamma^3_{44}&=&S\,\dot S\,(1+a^2+b^2)-a\,(a\,a_x+b\,b_x)\,S \nonumber \\
&&-b\,(a\,a_y+b\,b_y)\,S\ .\label{94d}
\end{eqnarray}
Hence the non--vanishing components of the shear tensor (\ref{94c}) are
\begin{eqnarray}
\sigma_{11}&=&-a_x\,S\ ,\ \sigma_{12}=\sigma_{21}=-a_y\,S\ ,\nonumber \\
\sigma_{14}&=&\sigma_{41}=(a\,a_x+b\,b_x)\,S\ ,\nonumber\\
\sigma_{22}&=&a_x\,S\ ,\ \sigma_{24}=\sigma_{42}=(a\,a_y+b\,b_y)\,S\ ,\nonumber\\
\sigma_{44}&=&-a\,(a\,a_x+b\,b_x)\,S-b\,(a\,a_y+b\,b_y)\,S\ .\label{94e}
\end{eqnarray}
We will make use of these components below. 

Next we turn to the Ricci tensor calculated with the metric given via the line element (\ref{93}). The non--vanishing components $R_{ij}$ of the Ricci tensor are found to be
\begin{eqnarray}
R_{11}&=&\frac{1}{2}\,S^2(\mu-p)-2\,a_x\,\dot S\ ,\label{r1}\\
R_{22}&=&\frac{1}{2}\,S^2(\mu-p)-2\,b_y\,\dot S\ ,\label{r2}\\
R_{12}&=&-2\,a_y\,\dot S\ ,\label{r3}\\
R_{33}&=&\mu+p\ ,\label{r4}\\
R_{34}&=&\frac{1}{2}\,S\,(\mu-p)\ ,\label{r5}\\
R_{14}&=&-\frac{1}{2}\,a\,S^2(\mu-p)+2\,(a\,a_x+b\,b_x)\,\dot S\ ,\label{r6}\\
R_{24}&=&-\frac{1}{2}\,b\,S^2(\mu-p)+2\,(a\,a_y+b\,b_y)\,\dot S\ ,\label{r7}\\
R_{44}&=&\frac{1}{2}\,S^2(\mu-p)\,(1+a^2+b^2)-2\,(a_x^2+a_y^2)\nonumber \\
&&-2\,a\,(a\,a_x+b\,b_x)\,\dot S-2\,b\,(a\,a_y+b\,b_y)\,\dot S,\label{r8}
\end{eqnarray}
with $\mu, p$ given by (\ref{33}). We can write (\ref{93}) in terms of basis 1--forms $\vartheta^{(a)}$, with $a=1, 2, 3, 4$ as
\begin{equation}\label{95}
ds^2=(\vartheta^{(1)})^2+(\vartheta^{(2)})^2+2\,\vartheta^{(3)}\,\vartheta^{(4)}=g_{(a)(b)}\,\vartheta^{(a)}\,\vartheta^{(b)}\ ,\end{equation}
with
\begin{eqnarray}
\vartheta^{(1)}&=&S\,(dx-a\,d\phi)=\vartheta_{(1)}=\vartheta_{(1)i}\,dx^i\ ,\label{96}\\
\vartheta^{(2)}&=&S\,(dy-b\,d\phi)=\vartheta_{(2)}=\vartheta_{(2)i}\,dx^i\ ,\label{97}\\
\vartheta^{(3)}&=&dT+\frac{S}{2}\,d\phi=\vartheta_{(4)}=\vartheta_{(4)i}\,dx^i\ ,\label{98}\\
\vartheta^{(4)}&=&S\,d\phi=\vartheta_{(3)}=\vartheta_{(3)i}\,dx^i\ .\label{99}
\end{eqnarray}
These 1--forms define a half null tetrad. The tetrad indices in brackets are lowered using the components of the metric tensor on the tetrad $g_{(a)(b)}$ so that $\vartheta_{(a)}=g_{(a)(b)}\,\vartheta^{(b)}$. Conversely $\vartheta^{(a)}=g^{(a)(b)}\,\vartheta_{(b)}$ with $g^{(a)(b)}$ given by $g^{(a)(b)}\,g_{(b)(c)}=\delta^{(a)}_{(c)}$ . The coordinate indices are raised with the inverse of the metric tensor $g^{ij}$ which satisfies $g^{ij}\,g_{jk}=\delta^i_k$. Hence we have from (\ref{96})--(\ref{99}):
\begin{eqnarray}
\vartheta^i_{(1)}\frac{\partial}{\partial x^i}&=&S^{-1}\frac{\partial}{\partial x}\ ,\ \vartheta^i_{(2)}\frac{\partial}{\partial x^i}=S^{-1}\frac{\partial}{\partial y}\ ,\ \vartheta^i_{(3)}\frac{\partial}{\partial x^i}=\frac{\partial}{\partial T}\ ,\nonumber\\
\vartheta^i_{(4)}\frac{\partial}{\partial x^i}&=&a\,S^{-1}\frac{\partial}{\partial x}+b\,S^{-1}\frac{\partial}{\partial y}-\frac{1}{2}\,\frac{\partial}{\partial T}+S^{-1}\frac{\partial}{\partial\phi}\ .\label{r9}
\end{eqnarray}
The non--vanishing components $R_{(a)(b)}=R_{ij}\,\vartheta^i_{(a)}\,\vartheta^j_{(b)}$ of the Ricci tensor (\ref{r1})--(\ref{r8}) are
\begin{eqnarray}
R_{(1)(1)}&=&\frac{1}{2}(\mu-p)-\frac{2\,\dot S}{S^2}\,a_x\ ,\label{100}\\
R_{(2)(2)}&=&\frac{1}{2}(\mu-p)-\frac{2\,\dot S}{S^2}\,b_y\ ,\label{101}\\
R_{(1)(2)}&=&-\frac{2\,\dot S}{S^2}\,a_y\ ,\label{102}\\
R_{(3)(3)}&=&\mu+p\ ,\label{103}\\
R_{(3)(4)}&=&-p\ ,\label{104}\\
R_{(4)(4)}&=&\frac{1}{4}(\mu+p)-\frac{2}{S^2}\,(a_x^2+a_y^2)\ .\label{105}
\end{eqnarray}
Also the tetrad components $u^{(a)}=u^i\,\vartheta^{(a)}_i$ of the 4--velocity (\ref{94a}) are 
\begin{eqnarray}
u^{(1)}&=&u_{(1)}=0\ ,\quad u^{(2)}=u_{(2)}=0\ ,\nonumber \\
u^{(3)}&=&u_{(4)}=\frac{1}{2}\ ,\quad u^{(4)}=u_{(3)}=-1\ .\label{x1}
\end{eqnarray}
When the Einstein tensor components $G_{(a)(b)}$ are decomposed with respect to this vector field it takes the algebraic form \cite{Ellis:1971}
\begin{eqnarray}
G_{(a)(b)}&=&\tilde\mu\,u_{(a)}\,u_{(b)}+\tilde p\,h_{(a)(b)}+q_{(a)}\,u_{(b)}\nonumber \\
&&+q_{(b)}\,u_{(a)}+\pi_{(a)(b)}\ ,\label{x2}
\end{eqnarray}
where $h_{(a)(b)}=g_{(a)(b)}+u_{(a)}\,u_{(b)}$, $q_{(a)}\,u^{(a)}=0$, $\pi_{(a)(b)}\,u^{(b)}=0$ and $\pi^{(a)}{}_{(a)}=0$. Since Einstein's field equations read $G_{(a)(b)}=T_{(a)(b)}$ the right hand side of (\ref{x2}) is the energy--momentum--stress tensor of the matter distribution. Here $\tilde\mu$ is the energy density measured by the observer with 4--velocity $u^{(a)}$ (equivalently $u^i$), $q_{(a)}$ is the energy or heat flux measured by this observer while $\tilde p$ and $\pi_{(a)(b)}=\pi_{(b)(a)}$ are the isotropic pressure and the anisotropic stress respectively of the matter distribution. These quantities can be obtained directly from the components of the Ricci tensor via the formulas \cite{Ellis:1971}
\begin{eqnarray}
R_{(a)(b)}\,u^{(a)}\,u^{(b)}&=&\frac{1}{2}(\tilde\mu+3\,\tilde p)\ ,\label{x3}\\
R_{(a)(b)}\,u^{(a)}\,h^{(b)}_{(c)}&=&-q_{(c)}\ ,\label{x4}\\
R_{(a)(b)}\,h^{(a)}_{(c)}\,h^{(b)}_{(d)}&=&\frac{1}{2}(\tilde\mu-\tilde p)\,h_{(c)(d)}+\pi_{(c)(d)}\ .\label{x5}
\end{eqnarray}
We first find that
\begin{equation}\label{106}
\tilde\mu=\mu-\frac{2}{S^2}\,(a_x^2+a_y^2)\ ,
\end{equation}
and
\begin{equation}\label{107}
\tilde p=p-\frac{2}{3\,S^2}\,(a_x^2+a_y^2)\ .
\end{equation}
The additional terms here, which are small second order perturbations when $a, b$ are considered small of first order, will be found to be due solely to the presence of gravitational waves. If we write $\tilde\mu=\mu+\mu_{W}$ and $\tilde p=p+p_{W}$ we see that $\mu_{W}=3\,p_{W}$. Continuing the calculation of $q_{(a)}$ and $\pi_{(a)(b)}$ using (\ref{x3}) and (\ref{x5}) we find that
\begin{equation}\label{108}
q_{(1)}=0\ ,\ q_{(2)}=0\ ,\ q_{(3)}=3\,p_W\ ,\ q_{(4)}=\frac{3}{2}\,p_W\ ,
\end{equation}
and $\pi_{(a)(b)}=0$ except for
\begin{eqnarray}
\pi_{(1)(1)}&=&-\frac{2\,\dot S}{S^2}\,a_x-p_W\ ,\label{109}\\
\pi_{(1)(2)}&=&-\frac{2\,\dot S}{S^2}\,a_y\ ,\label{110}\\
\pi_{(2)(2)}&=&\frac{2\,\dot S}{S^2}\,a_x-p_W\ ,\label{111}\\
\pi_{(3)(3)}&=&2\,p_W\ ,\label{112}\\
\pi_{(3)(4)}&=&p_W\ ,\label{113}\\
\pi_{(4)(4)}&=&\frac{1}{2}\,p_W\ .\label{114}
\end{eqnarray}
Substituting these results into the energy--momentum--stress tensor $T_{(a)(b)}$ given by the right hand side of (\ref{x2}) above we arrive at
\begin{equation}\label{114'}
T_{(a)(b)}=\mu\,u_{(a)}\,u_{(b)}+p\,h_{(a)(b)}+\Pi_{(a)(b)}+\Lambda_{(a)(b)}\ ,
\end{equation}
with $\Pi_{(a)(b)}$ vanishing except for
\begin{eqnarray}
\Pi_{(1)(1)}&=&-\frac{2\,\dot S}{S^2}\,a_x\ ,\ \Pi_{(1)(2)}=\Pi_{(2)(1)}=-\frac{2\,\dot S}{S^2}\,a_y\ ,\nonumber \\
\Pi_{(2)(2)}&=&-\frac{2\,\dot S}{S^2}\,b_y\ ,\label{114''}
\end{eqnarray}
and with $\Lambda_{(a)(b)}$ vanishing except for
\begin{eqnarray}
&&\Lambda_{(4)(4)}=\mu_W\ \ \Rightarrow\ \nonumber \\
&&\Lambda_{ij}=S^2\mu_W\,\phi_{,i}\,\phi_{,j}=-2\,(a_x^2+a_y^2)\,\phi_{,i}\,\phi_{,j}\ .\label{114'''}
\end{eqnarray}
The coordinate components of the energy--momentum--stress tensor read
\begin{equation}\label{114a}
T_{ij}=\mu\,u_i\,u_j+p\,h_{ij}+\Pi_{ij}+\Lambda_{ij}\ ,
\end{equation}
with the non--vanishing components $\Pi_{ij}$ given by
\begin{eqnarray}
\Pi_{11}&=&-2\,a_x\,\dot S\ ,\ \Pi_{12}=\Pi_{21}=-2\,a_y\,\dot S\ ,\nonumber \\
\Pi_{14}&=&\Pi_{41}=2\,(a\,a_x+b\,b_x)\,\dot S\ ,\nonumber\\
\Pi_{22}&=&-2\,b_y\,\dot S\ ,\ \Pi_{24}=\Pi_{42}=2\,(a\,a_y+b\,b_y)\,\dot S\ ,\nonumber\\
\Pi_{44}&=&-2\,\{a\,(a\,a_x+b\,b_x)+b\,(a\,a_y+b\,b_y)\}\,\dot S\ ,\label{114b}
\end{eqnarray}
from which it follows that $g^{ij}\,\Pi_{ij}=0=\Pi_{ij}\,u^j$. We thus see that the matter distribution consists of the isotropic perfect fluid of the cosmological model, an anisotropic stress and, as evident from (\ref{114'''}), matter travelling with the speed of light accompanying the gravitational radiation which is present for the following reason: The non--vanishing components $C_{ijkl}$ of the Weyl conformal curvature tensor are 
\begin{eqnarray}
C_{1414}&=&-S^2\{a'_x+a\,a_{xx}+b\,b_{xx}\}\ ,\label{r10}\\
C_{2424}&=&-S^2\{b'_y+a\,a_{yy}+b\,b_{yy}\}=-C_{1414}\ ,\label{r11}\\
C_{1424}&=&-S^2\{a'_y+a\,b_{xx}+b\,a_{yy}\}\ .\label{r12}
\end{eqnarray}
The non--vanishing independent tetrad components of the Weyl tensor are $C_{(1)(4)(1)(4)}$ and $C_{(1)(4)(2)(4)}$. Thus there is only one non--vanishing complex Newman--Penrose \cite{Newman:Penrose:1962} component of this tensor, namely,
\begin{eqnarray}\label{115}
\Psi_4&=&-C_{(1)(4)(1)(4)}+i\,C_{(1)(4)(2)(4)} \nonumber \\
&=&\frac{(a'_x-i\,a'_y)}{S^2}+\frac{(a+i\,b)(a_x-i\,a_y)_x}{S^2}\ .
\end{eqnarray}
This means that the gravitational field is purely radiative (Type N in the Petrov classification) with propagation direction $\phi_{,i}$ in space--time.

The tetrad components $\sigma_{(a)(b)}$ of the shear  tensor associated with the matter world lines are obtained from $\sigma_{(a)(b)}=\sigma_{ij}\,\vartheta^i_{(a)}\,\vartheta^j_{(b)}$ using (\ref{94e}). We find that the non--vanishing tetrad components are
\begin{eqnarray}\label{116'}
\sigma_{(1)(1)}&=&-a_x\,S^{-1}\ ,\ \sigma_{(1)(2)}=\sigma_{(2)(1)}=-a_y\,S^{-1}\ ,\nonumber \\
\sigma_{(2)(2)}&=&-b_y\,S^{-1}\ .
\end{eqnarray}
Comparing these with $\Pi_{(a)(b)}$ given in (\ref{114''}) we have
\begin{equation}\label{117'}
\Pi_{(a)(b)}=\frac{2\,\dot S}{S}\,\sigma_{(a)(b)}=S\,\phi_{,i}{}^{;i}\,\sigma_{(a)(b)}\ ,
\end{equation}
with $\phi_{,i}{}^{;i}$ given by (\ref{24}). This is an \emph{exact} version of the approximate equation (\ref{91}) in this $k=0$ case.

\section{An Exact Model When $k=-1$}\label{sec:7}

Consider now the line element, in coordinates $x^i=(x, y, T, \phi)$ for $i=1, 2, 3, 4$,
\begin{eqnarray}
ds^2&=&g_{ij}\,dx^i\,dx^j \nonumber \\
&=&S^2\,e^{2(\phi+\tau)}\{(dx-a\,d\phi)^2+(dy-b\,d\phi)^2\}\nonumber \\
&&+2\,S\,dT\,d\phi+S^2d\phi^2\ ,\label{116}
\end{eqnarray}
with $S=S(T), \tau=\tau(T), a=a(x, y, \phi), b=b(x, y, \phi)$ and
\begin{equation}\label{117}
\frac{d\tau}{dT}=\frac{1}{S}, \ a_x=-b_y\ \ \ {\rm and}\ \ \ a_y=b_x\ ,
\end{equation}
with the subscripts denoting partial derivatives as before. The coordinates $x, y$ here differ from the coordinates $x, y$ in (\ref{48}) by terms which are small of first order, when $a, b$ are small of first order and second order and smaller terms are neglected.

Here the matter 4--velocity is again given by (\ref{94a}) and the integral curves of this unit timelike vector field are geodesic, twist--free, expanding with expansion $\theta=u^i{}_{;i}=3\,\dot S/S$ and with non--vanishing shear tensor given by (\ref{94c}) but in this case the non--vanishing Christoffel symbols (\ref{94d}) are replaced by
\begin{eqnarray}
\Gamma^3_{11}&=&(\dot S-a_x)S\,e^{2(\phi+\tau)}\ ,\ \Gamma^3_{22}=(\dot S-b_y)Se^{2(\phi+\tau)}\ ,\nonumber \\
\Gamma^3_{12}&=&-a_y\,S\,e^{2(\phi+\tau)}\ , \Gamma^3_{33}=S^{-1}\dot S\ ,\ \nonumber\\
\Gamma^3_{14}&=&(-a\,\dot S+a\,a_x+b\,b_x)S\,e^{2(\phi+\tau)}\ ,\ \Gamma^3_{34}=\dot S\ ,\nonumber\\
\Gamma^3_{24}&=&(-b\,\dot S+a\,a_y+b\,b_y)S\,e^{2(\phi+\tau)}\ , \nonumber \\
\Gamma^3_{44}&=&S\,\dot S\{1+(a^2+b^2)e^{2(\phi+\tau)}\}\nonumber\\
&-&\{a(a\,a_x+b\,b_x)+b(a\,a_y+b\,b_y)\}S\,e^{2(\phi+\tau)}\ .\label{r12b}
\end{eqnarray}
With these expressions the surviving components $\sigma_{ij}$ of the shear tensor are
\begin{eqnarray}
\sigma_{11}&=&-a_x\,S\,e^{2(\phi+\tau)}\ ,\ \sigma_{12}=-a_y\,S\,e^{2(\phi+\tau)}\ ,\nonumber \\
\sigma_{14}&=&(a\,a_x+b\,b_x)S\,e^{2(\phi+\tau)}\ ,\nonumber\\
\sigma_{22}&=&a_x\,S\,e^{2(\phi+\tau)}\ ,\ \sigma_{24}=(a\,a_y+b\,b_y)\,S\,e^{2(\phi+\tau)}\ ,\nonumber\\
\sigma_{44}&=&-a\,(a\,a_x+b\,b_x)\,S\,e^{2(\phi+\tau)} \nonumber \\
&&-b\,(a\,a_y+b\,b_y)\,S\,e^{2(\phi+\tau)}\ .\label{r13}
\end{eqnarray}

Next we shall require the non--vanishing components $R_{ij}$ of the Ricci tensor calculated with the metric given by the line element (\ref{116}). These are found to be
\begin{eqnarray}
R_{11}&=&\left (\frac{1}{2}S^2(\mu-p)-2\,a_x(\dot S+1)\right )e^{2(\phi+\tau)}\ ,\label{r14}\\
R_{22}&=&\left (\frac{1}{2}S^2(\mu-p)-2\,b_y(\dot S+1)\right )e^{2(\phi+\tau)}\ ,\label{r15}\\
R_{12}&=&-2\,a_y\,(\dot S+1)e^{2(\phi+\tau)}\ ,\label{r16}\\
R_{33}&=&\mu+p\ ,\label{r17}\\
R_{34}&=&\frac{1}{2}S\,(\mu-p)\ ,\label{r118}\\
R_{14}&=&\left (-\frac{1}{2}\,a\,S^2(\mu-p)+2\,(\dot S+1)(a\,a_x+b\,b_x)\right )e^{2(\phi+\tau)},\nonumber \\
\label{r119}\\
R_{24}&=&\left (-\frac{1}{2}\,b\,S^2(\mu-p)+2\,(\dot S+1)(a\,a_y+b\,b_y)\right )e^{2(\phi+\tau)},\nonumber \\
\label{r120}\\
R_{44}&=&-2\,(a_x^2+a_y^2)+\frac{1}{2}\,S^2(\mu-p)\{1+(a^2+b^2)\,e^{2(\phi+\tau)}\}\nonumber\\
&&-2\{a\,(a\,a_x+b\,b_x)+b\,(a\,a_y+b\,b_y)\}(\dot S+1)e^{2(\phi+\tau)}.\nonumber \\
\label{r121}
\end{eqnarray}
The proper density $\mu$ and isotropic pressure $p$ appearing here are given in (\ref{64}).

Writing the line element (\ref{116}) in the form (\ref{95}) the basis 1--forms are now
\begin{eqnarray}
\vartheta^{(1)}&=&S\,e^{\phi+\tau}(dx-a\,d\phi)=\vartheta_{(1)}=\vartheta_{(1)i}\,dx^i\ ,\label{r122}\\
\vartheta^{(2)}&=&S\,e^{\phi+\tau}(dy-b\,d\phi)=\vartheta_{(2)}=\vartheta_{(2)i}\,dx^i\ ,\label{r123}\\
\vartheta^{(3)}&=&dT+\frac{1}{2}\,S\,d\phi=\vartheta_{(4)}=\vartheta_{(4)i}\,dx^i\ ,\label{r124}\\
\vartheta^{(4)}&=&S\,d\phi=\vartheta_{(3)}=\vartheta_{(3)i}\,dx^i\ .\label{r125}\end{eqnarray}
From these and the metric given by (\ref{116}) we have
\begin{eqnarray}
\vartheta^i_{(1)}\frac{\partial}{\partial x^i}&=&S^{-1}e^{-\phi-\tau}\frac{\partial}{\partial x}\ ,\nonumber \\
\vartheta^i_{(2)}\frac{\partial}{\partial x^i}&=&S^{-1}e^{-\phi-\tau}\frac{\partial}{\partial y}\ ,\nonumber \\
\vartheta^i_{(3)}\frac{\partial}{\partial x^i}&=&\frac{\partial}{\partial T}\ ,\nonumber\\
\vartheta^i_{(4)}&=&a\,S^{-1}\frac{\partial}{\partial x}+b\,S^{-1}\frac{\partial}{\partial y} \nonumber \\
&&-\frac{1}{2}\,\frac{\partial}{\partial T}+S^{-1}\frac{\partial}{\partial\phi}\ .\label{r126}
\end{eqnarray}
Now the non--zero tetrad components $R_{(a)(b)}=R_{ij}\,\vartheta^i_{(a)}\,\vartheta^j_{(b)}$ of the Ricci tensor (\ref{r14})--(\ref{r121}) are
\begin{eqnarray}
R_{(1)(1)}&=&\frac{1}{2}(\mu-p)-\frac{2\,(\dot S+1)}{S^2}\,a_x\ ,\label{r127}\\
R_{(2)(2)}&=&\frac{1}{2}(\mu-p)-\frac{2\,(\dot S+1)}{S^2}\,b_y\ ,\label{r128}\\
R_{(1)(2)}&=&-\frac{2\,(\dot S+1)}{S^2}\,a_y\ ,\label{r129}\\
R_{(3)(3)}&=&\mu+p\ ,\label{r130}\\
R_{(3)(4)}&=&-p\ ,\label{r131}\\
R_{(4)(4)}&=&\frac{1}{4}(\mu+p)-\frac{2}{S^2}\,(a_x^2+a_y^2)\ .\label{r132}
\end{eqnarray}
As mentioned above the 4--velocity of matter is again given by (\ref{94a}). Its tetrad components $u^{(a)}$ are given by (\ref{x1}) in the present case also. Now reading $\tilde\mu, \tilde p, q_{(a)}, \pi_{(a)(b)}$ from (\ref{x3})--(\ref{x5}) with $R_{(a)(b)}$ given now by (\ref{r127})--(\ref{r132}) we find that
\begin{equation}
\tilde\mu=\mu+\mu_W\ ,\quad \tilde p=p+p_W\ \nonumber
\end{equation}
with
\begin{equation}
\mu_W=3\,p_W=-\frac{2}{S^2}\,(a_x^2+a_y^2)\ ,\label{r133}
\end{equation}
\begin{equation}\label{r134}
q_{(1)}=q_{(2)}=0\ ,\ q_{(3)}=\mu_W\ ,\ q_{(4)}=\frac{1}{2}\mu_W\ ,
\end{equation}
and the non--vanishing tetrad components $\pi_{(a)(b)}$ of the anisotropic stress are
\begin{eqnarray}
\pi_{(1)(1)}&=&-\frac{2\,(\dot S+1)}{S^2}\,a_x-p_W\ ,\label{s1}\\
\pi_{(1)(2)}&=&-\frac{2\,(\dot S+1)}{S^2}\,a_y\ ,\label{s2}\\
\pi_{(2)(2)}&=&\frac{2\,(\dot S+1)}{S^2}\,a_x-p_W\ ,\label{s3}\\
\pi_{(3)(3)}&=&2\,p_W\ ,\label{s4}\\
\pi_{(3)(4)}&=&p_W\ ,\label{s5}\\
\pi_{(4)(4)}&=&\frac{1}{2}\,p_W\ .\label{s6}\end{eqnarray}
Substituting these into the energy--momentum--stress tensor $T_{(a)(b)}$ given by the right hand side of (\ref{x2}) we arrive at the form of energy--momentum--stress tensor given in (\ref{114'}). In this case the non--vanishing $\Lambda_{(a)(b)}$ and $\Pi_{(a)(b)}$ are given by $\Lambda_{(4)(4)}=\mu_W$ and 
\begin{eqnarray}\label{s7}
\Pi_{(1)(1)}&=&-\frac{2\,(\dot S+1)}{S^2}\,a_x\ ,\ \Pi_{(1)(2)}=-\frac{2\,(\dot S+1)}{S^2}\,a_y\ ,\nonumber \\
\Pi_{(2)(2)}&=&-\frac{2\,(\dot S+1)}{S^2}b_y\ .
\end{eqnarray}
The coordinate components of the energy--momentum--stress tensor again have the algebraic form (\ref{114a}) with 
$\Lambda_{ij}=\mu_W\,S^2\phi_{,i}\,\phi_{,j}$ and the non--vanishing components $\Pi_{ij}$ are given by
\begin{eqnarray}
\Pi_{11}&=&-2\,a_x\,(\dot S+1)\,e^{2(\phi+\tau)}\ ,\nonumber \\
\Pi_{12}&=&\Pi_{21}=-2\,a_y\,(\dot S+1)\,e^{2(\phi+\tau)}\ ,\nonumber\\
\Pi_{14}&=&\Pi_{41}=2\,(a\,a_x+b\,b_x)\,(\dot S+1)\,e^{2(\phi+\tau)}\ ,\nonumber \\
\Pi_{22}&=&-2\,b_y\,(\dot S+1)\,e^{2(\phi+\tau)}\ ,\nonumber\\
\Pi_{24}&=&\Pi_{42}=2\,(a\,a_y+b\,b_y)\,(\dot S+1)\,e^{2(\phi+\tau)} ,\nonumber\\
\Pi_{44}&=&-2\,\{a\,(a\,a_x+b\,b_x)+b\,(a\,a_y+b\,b_y)\}\nonumber \\
&&\times(\dot S+1)\,e^{2(\phi+\tau)}\ . \label{s7'}
\end{eqnarray}
Hence the matter distribution in this case is again composed of the cosmological perfect fluid supplemented with an anisotropic stress $\Pi_{(a)(b)}$ and matter travelling with the gravitational waves. The non--vanishing coordinate components $C_{ijkl}$ of the Weyl tensor in this case are
\begin{eqnarray}
C_{1414}&=&-S^2\{a_x+a'_x+a\,a_{xx}+b\,b_{xx}\}e^{2(\phi+\tau)}\nonumber \\
&=&-C_{2424}\ ,\label{s8}\\
C_{1424}&=&-S^2\{b_x+a'_y+a\,b_{xx}+b\,a_{yy}\}e^{2(\phi+\tau)}\ .\label{s9}
\end{eqnarray}
The corresponding tetrad components yield the same expression for $\Psi_4$ given in (\ref{115}) confirming that here we are again dealing with a purely radiative Weyl tensor.

In this case the shear tensor (\ref{r13}) has non--vanishing tetrad components
\begin{eqnarray}
\sigma_{(1)(1)}&=&-S^{-1}a_x\ ,\nonumber \\
\sigma_{(1)(2)}&=&-S^{-1}a_y\ ,\ \sigma_{(2)(2)}=-S^{-1}a_y\ ,\label{s10}
\end{eqnarray} 
and thus it follows from (\ref{s7}) that
\begin{equation}\label{s11}
\Pi_{(a)(b)}=\frac{2\,(\dot S+1)}{S}\,\sigma_{(a)(b)}=S\,\phi^{,i}{}_{;i}\,\sigma_{(a)(b)}\ .
\end{equation}
This is an \emph{exact} version of the approximate equation (\ref{91}) in this $k=-1$ case.
 
\section{Discussion}\label{sec:discussion}

We have derived gravitational wave perturbations of the isotropic universes corresponding to $k=0$ and $k=-1$ in sections \ref{sec:3} and \ref{sec:4}. In sections \ref{sec:6} and \ref{sec:7} we have constructed exact models of gravitational waves in these isotropic universes which have the property that if the gravitational wave variables (the functions $a, b$ appearing in (\ref{93}) and (\ref{116}) and their derivatives) are small of first order, and if second order and smaller terms are neglected, then these exact models reproduce the perturbative models described in sections \ref{sec:3} and \ref{sec:4}. A striking property of the exact models is how similar the resulting matter distribution resembles the matter distribution in the perturbative cases. The matter distribution of the former differs qualitatively from that of the latter simply by the appearance of lightlike matter traveling with the gravitational waves. In the perturbative scenario this lightlike matter is small of second order and thus is neglected. It is not unusual in explicit examples to find gravitational radiation accompanied by matter travelling with the speed of light \cite{Hogan:Puetzfeld:2021}. In the exact models above the presence of lightlike matter is of great importance since if it is not present then the functions $a, b$ are independent of $x, y$ and can be disposed of. In this case the models become simply the $k=0$ or $k=-1$ isotropic cosmologies.

\begin{acknowledgments}
This work was funded by the Deutsche Forschungsgemeinschaft (DFG, German Research Foundation) through the grant PU 461/1-2 -- project number 369402949 (D.P.). 
\end{acknowledgments}

\bibliographystyle{unsrtnat}
\bibliography{isocosm_bibliography}
\end{document}